# Lateral diffusion of phospholipids in artificial cell membranes measured by single shallow NV centers.

*Farida Shagieva\*,‡, Andrea Zappe‡, Daniel Cohen†, Andrej Denisenko‡, Alex Retzker† and Jörg Wrachtrup‡*

‡Institute for Quantum Science and Technology (IQST), University of Stuttgart, 3rd Institute of Physics, Stuttgart, Germany.
† Racah Institute of Physics, The Hebrew University of Jerusalem, Jerusalem, 91904, Givat Ram, Israel.

**ABSTRACT.** We measure diffusion of organic molecules located a few nanometers from the diamond surface. To study molecular diffusion, we perform local detection of nuclear magnetic resonance, based on single shallow Nitrogen-Vacancy (NV) centers in diamond. Specifically, we demonstrate measurements of translational diffusion coefficient of phospholipids in an artificial cell membrane by employing correlation spectroscopy. An analysis of correlation decay curves using different diffusion models shows, that a simple 2D diffusion model gives satisfactory diffusion coefficients, although the choice of the model affects the extracted numbers. We find significant differences among the diffusion coefficients measured by different single NV centers, which we attribute to local heterogeneities of the lipid layers, likely caused by the supporting diamond substrate.

The cell membrane is a site of important biological processes[1]. It is made of a fluid phospholipid bilayer with embedded proteins and carbohydrates[2], which are continuously moving within the membrane. This motion has numerous biological roles (e.g. cell signaling and membrane transport). Hence, lipid mobility has been extensively investigated in recent years in both cellular and model bilayers. Lipid lateral diffusion, the translational diffusion along the layer of phospholipids, is a fundamental process exploited by cells e.g. to enable complex protein structural and dynamic reorganizations. There are many mechanisms constraining lateral diffusion[3,4], which result in nanoscale heterogeneity of phospholipid mobility. One of them is the presence of the so-called lipid rafts: heterogeneous, dynamic membrane nanodomains (10–200 nm)[5,6]. Measuring molecular mobility in plasma membranes on the nanoscale is thus key to understanding the functioning of various cellular processes. This is why numerous attempts have been made to study the dynamics not only of natural cell membranes but also of model systems like phospholipid bilayers and monolayers.

The existing methodologies to study the diffusion in plasma membranes on the nanoscale rely mostly on fluorescent labels[7,8,9,10]. The attached nanoparticles and fluorescent tags, such as fluorophores, are often almost the size of the lipid molecules, which inevitably alters the native behaviour of lipids and the diffusion rates because of unspecific interactions with other membrane components.

A method which would enable the label-free studying of the diffusion heterogeneities on the nanoscale is highly desirable. During the last decade NV center in diamond have made impact on biosensing owing to a vast scope of measurable quantities, low cytotoxicity and different measurement configurations[11,12,13,14,15]. Early studies of dynamical processes in artificial cell membranes were made by performing diffusion-mediated relaxometry of nanodiamonds incorporated into the Gadolinium-labeled phospholipids' bilayer[16]. Another approach, which could eliminate labeling and enable completely non-invasive diffusion studies, is based on the recording of temporal correlations among the nuclear spins measured by nanoscale nuclear magnetic resonance (NMR) using single NV centers in diamond[17,18,19]. The technique allows studying the dynamics of small nuclear spin ensembles within the nanoscopic sample volume sensed by the NV center. NV NMR spectroscopy does not require fluorescent labeling and measurements are performed at room temperature. Besides, only a few 10 nm large sample space is measured. Thus, correlation spectroscopy with NV centers can provide a high degree of localization accuracy necessary to study heterogeneity of lateral diffusion of phospholipids on the nanoscale. It has been shown that correlation spectroscopy with a single shallow NV center allows studying nuclear spin dynamics in homogeneous solid and liquid samples[27]. Here we exploit correlation spectroscopy with single shallow NV centers to study the diffusion of phospholipids in artificial cell membranes, formed by a monolayer of phospholipids, which is the first step towards the investigation of non-uniformity of diffusion in spatially contiguous lipid clusters.

As an object that mimics the cell membrane, we used micelles dissolved in heavy water (Figure S1). These are spherical nanoparticles with a characteristic size of 100-150 nm in diameter formed by a monolayer of phospholipids, in which the inner volume is filled by perfluoro-crown-ether. The thickness of such a monolayer is around 3-4 nm[20], which lies within the sensitivity range of a shallow NV center.

To perform the experiments, we have used approximately 30 μm thick diamond membranes implanted with single shallow NV centers and textured with nanopillars[21]. Pillar-shaped waveguides allow achieving a net photon flux up to ~ $1.7 \times 10^6$ s$^{-1}$, which significantly decreases the signal acquisition time and can result in up to 5 times better sensitivity. In the current work, the experiments were done at a count rate of 500-700 kcounts/s. All results have been acquired at room temperature (20°C).

A droplet of the micelle-containing solution was placed on the diamond surface, forming a configuration presented in Figure 1a. The NV center inside the [100] oriented diamond is located several nanometers below the surface. Due to the fast isotope exchange reaction[22,23,24] between heavy water and the adsorbed water layer on the diamond interface, we conclude that the hydrogen signal originates from the phospholipid molecules (length of the hydrocarbon chain 2 nm and 1.5 nm of the hydrophilic part[20]). The $^{31}$P nuclear spins are situated in the center of the headgroup with the typical area of 0.71 nm$^2$ and 1.5 nm thickness[25,26]. The fluorine atoms are located only inside the micelles in the ether. Knowing the number of atoms per unit volume, the following nuclear densities can be obtained: $\rho_{1H} = 3.5 \times 10^{28}$ m$^{-3}$, $\rho_{19F} = 2.7 \times 10^{28}$ m$^{-3}$, $\rho_{31P} = 9.4 \times 10^{26}$ m$^{-3}$. $^{13}$C nuclear spins were not taken into account, because their signal cannot be isolated from intrinsic carbon in the diamond lattice.

Thus, in this system, there are three kinds of nuclei ($^1$H, $^{31}$P in phospholipids and $^{19}$F in the ether), which can generate an NMR signal. However, diffusion can be measured only for those nuclear spins whose signal can be detected. For this reason, we have calculated the root-mean-square magnetic field $B_{rms}$ created by a layer of nuclear spins on the diamond surface at the NV center spin cite as a function of NV depth[27]:

$$B_{rms}^2 = \frac{5\pi}{24}\left(\frac{\mu_0}{4\pi}\right)^2 \mu_n^2 \rho_n \left[\frac{1}{d^3} - \frac{1}{(d+h)^3}\right] \quad (1)$$

where $\mu_0$ and $\mu_n$ are vacuum magnetic permeability and nuclear magnetic moment, $\rho_n$ is the nuclear spin density, d and h are NV depth and thickness of a nuclei-containing layer (3.5 nm for $^1$H, 1.5 nm for $^{31}$P and 15 nm for $^{19}$F), correspondingly. For calculating the $B_{rms}$ from fluorine nuclei we added the thickness of the phospholipids monolayer (3.5 nm) to the NV depth.

Taking into account the average relaxation times of shallow NV centers in the used diamond membrane ($T_2 \sim 12$ μs, $T_1 \sim 200$ μs), the simulation results depicted in Figure 1b clearly indicate that only the hydrogen signal is strong enough to be detected. Therefore, we have focused on $^1$H correlation spectroscopy. An example of the hydrogen signal from micelles detected via dynamical decoupling (DD) noise spectroscopy is presented in Figure 1c. On the one hand, the contrast of the nano-NMR signal is proportional to the phase, acquired by NV center during the application of the DD sequence. Therefore, the larger is the number of repetitions N, the higher is the contrast of the NMR signal. On the other hand, one cannot infinitely increase N because of the finite coherence time of the NV center's electron spin. Therefore, we optimized the number of repetitions until the best signal contrast was obtained. The $B_{rms}$ of the acquired signals are depicted by circles in Figure 1b, where the NV depth was determined by immersion oil measurements[28,29,30,31].

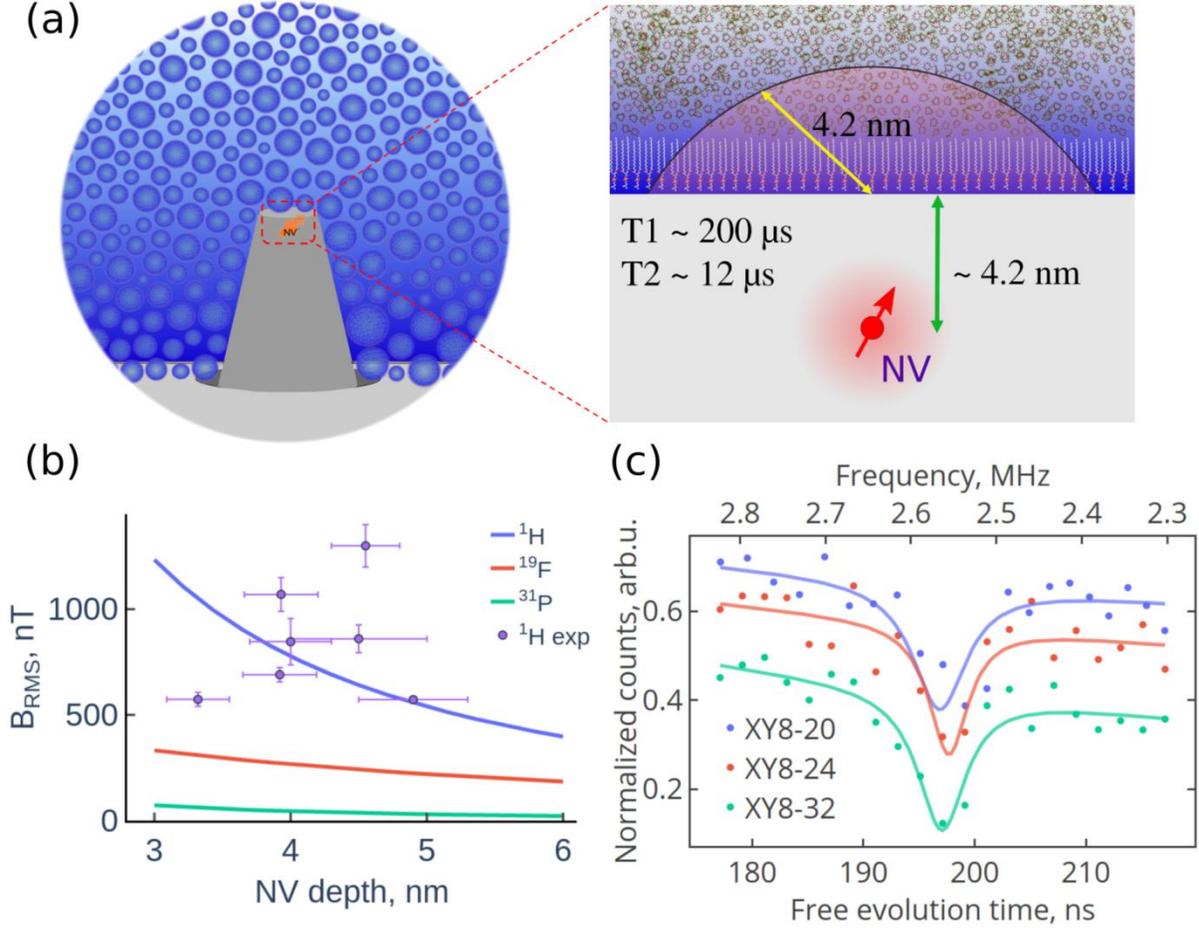

**Figure 1.** Experimental system. (a) Diamond membrane with nanopillars covered by micelles and the model used for $B_{rms}$ calculations. (b) $B_{rms}$ generated by nuclear spins in the micelles-containing solution calculated using Equation 1. Dots with error bars show the experimental data for the hydrogen signal strength in phospholipids. The NV depth was determined by the strength of the NMR signal from immersion oil (Fluka Analytical 10976) measurements with hydrogen nuclear density ($\rho_{1H} = 5 \times 10^{28}$ m$^{-3}$). (c) The hydrogen signal detected with the same NV center, applying a different number of repetitions of the XY8 DD sequence at 595.7 G.

The microwave pulse sequence of the correlation protocol consists of two DD pulse sequences separated by the intersequence time $\tilde{\tau}$ (Figure 2a). The detected signal represents the correlations between the phases (generated by i-th spin in j-th molecule) acquired by NV center within the first and the second XY8-N pulse trains[25,26,27]:

$$S(\tau, \tilde{\tau}) \sim \sum_{i,j} \langle \varphi_{i,j}(0), \varphi_{i,j}(\tilde{\tau}) \rangle \sim \cos(2\pi\omega_L(2(N \times 8)\tau + \tilde{\tau})) e^{-\frac{2(N \times 8)\tau}{T_2^{NV}}} \sum_{i,j} p_{i,j}(\tilde{\tau}) e^{-\frac{\tilde{\tau}}{T_2^{i,j}}} e^{-\frac{\tilde{\tau}}{T_1^{NV}}}, \quad (2)$$

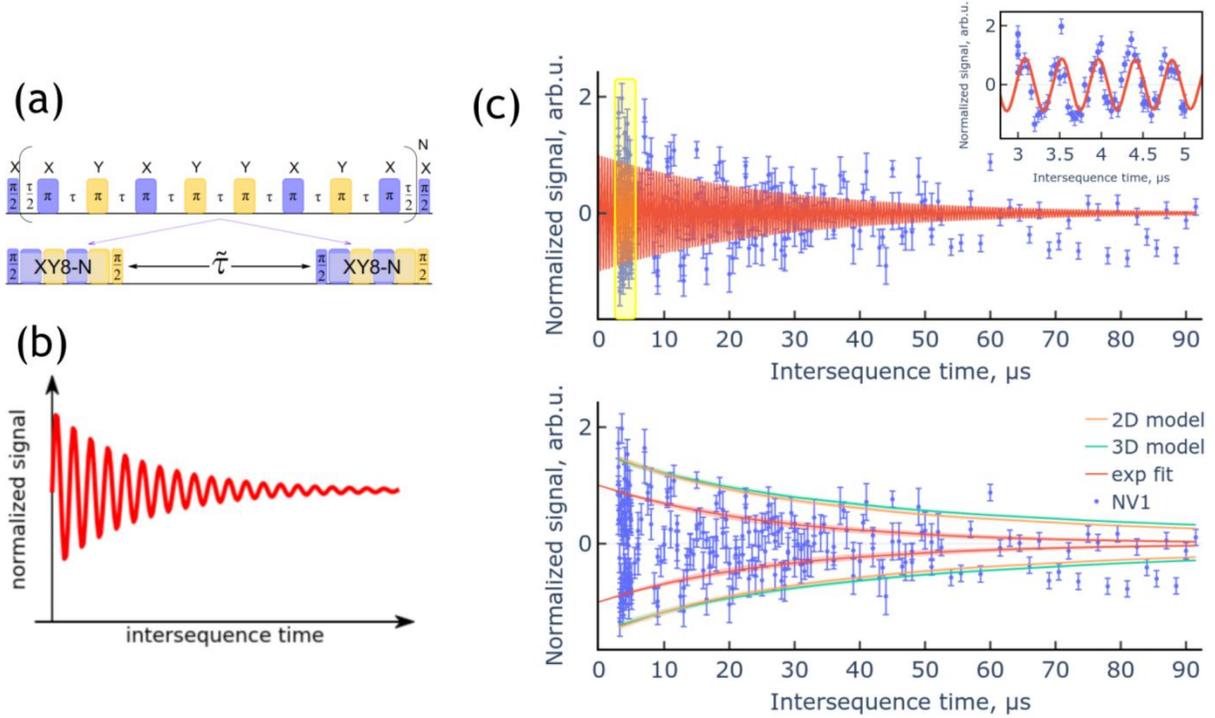

**Figure 2.** Correlation measurements. (a) Pulse sequence of the correlation protocol consists of two DD sequences separated by an intersequence time $\tilde{\tau}$. (b) The resulting signal has the shape of decaying cosine and is proportional to the correlation of phases, acquired by NV center during the first and second application of the XY8-N sequences. (c) Experimental data obtained with XY8-10 at 532 G applied to 3.93 nm deep NV center fitted by a decaying cosine (inset) and the same data where the fit envelopes are shown (main figure). The inset shows oscillations with a frequency of 2.2689 ± 0.0009 MHz ($^1$H Larmor frequency 2.13 MHz), which are observed at short time scales (range corresponds to data in the yellow rectangle).

where the first multiplier corresponds to the oscillations with the Larmor frequency $\omega_L$ of the nuclear spins, the second is the amplitude factor and the expression under the sum sign describes the envelope of the signal. The resulting signal has the shape of an exponentially decaying cosine (Figure 2b). Recently, it has been demonstrated that the asymptotic behavior of the correlation function can deviate from the exponential decay model for different geometries[32,33,34]. Namely, for times $\tilde{\tau}$ exceeding the characteristic diffusion time $\tau_D = d^2/D$ the correlation decays as a power-law:

$$C(t) \propto \cos(\widetilde{\omega}\tilde{\tau})\tilde{\tau}^{-\alpha}. \quad (3)$$

For the case of particles diffusing in a three-dimensional half-space $\alpha = 3/2$, whereas for a two-dimensional layer $\alpha = 1$. Therefore, the analysis of the envelope of the correlation signal is of particular interest, because it helps to understand the detection method and diffusion process in detail.

This power-law scaling of the correlation function can be understood as follows. The FT of the correlation function, goes as $PS(\omega = 0) \propto B_{rms}^2 \tau_D \propto d^{-1}$ (see Equation 1 and definition of $\tau_D$ above) by dimensional analysis, which reflects the NV's effective interaction region of radius $\sim d$. Nuclei that are close to (far from) the diamond surface induce high (low) frequency noise at the NV's location, because small changes in their position cause large (minute) fluctuations in the magnetic field. The DD sequence creates an effective cut-off scale $l = \sqrt{D/\omega}$ for the high frequency noise, such that the measured signal originates from nuclei found outside of a hemisphere of radius $\sim l$. By this argument $PS(\omega) \propto d^{-1} - l^{-1} = d^{-1} - \sqrt{\omega/D}$ and its inverse FT implies a $\tilde{\tau}^{-3/2}$ scaling for the correlation function.

Having the NMR signal with the optimized number of pulses N, we performed correlation spectroscopy with N/2 repetitions and intersequence time $\tilde{\tau}$ between pulse trains. The information about the molecular mobility is contained in the envelope of the correlation signal and is determined by the expression under the sum in Equation 2. Therefore, the measurement procedure was as follows.

First, we performed measurements on short time scales $\tilde{\tau}$ = 3-5 μs (Figure 2c, inset) to see the expected oscillations with the frequency close to Larmor frequency of hydrogen nuclear spin in accordance with Equation 2. After that, the intersequence time $\tilde{\tau}$ was increased until the decay of the signal could be observed. In this case, the number of measurement points was decreased to reduce the total measurement time so that the distance between the measurement points was larger than the Larmor period of $^1$H nuclear spins. Then, the experimental data taken at different ranges were combined and fitted by the decaying cosine function $S(t) = S_0 \cos(\widetilde{\omega} t) e^{-\alpha t}$, 2D- and 3D-models according to Equation 3. The 2D-model corresponds to the Brownian motion of the nuclear spins within the layer of thickness 3.5 nm and 3D-model describes the diffusion within the half-space. If one of these models yields a better fit to the experimental data, it allows to differentiate between a single layer or a multilayer of phospholipids. Figure 2d presents an example of the correlation signal fitted by exponentially decaying cosine together with the same data with different fit envelopes. For the available dataset (Figure S3) we can conclude that the data quality does not allow to differentiate between the two cases described above. Partially, this is caused by the small depth of the used NV centers, i.e. a monolayer of phospholipids already covers most of the detection volume and no signal from a second layer is expected.

The contrast of the signal depends on the amplitude factor $\exp(-2(N \times 8)\tau/T_2^{NV})$ in Equation 2. Taking into account that the average $T_2^{NV}$ time for this sample is ~ 12 μs, the expected contrast cannot be more than 5%. In the experiment, the maximal observed contrast of the detected correlations was 3.9%.

When we analyze the experimental data using Equation 3, the diffusion coefficient is the fit parameter, because it directly defines the conditional probability occurring in the correlation measurements. To extract the coefficient of translational diffusion $D_T$ in the case of exponential decay, it is convenient to rewrite the correlation signal in the following way[27]:

$$\sum_j p_j(\tilde{\tau}) e^{-\frac{\tilde{\tau}}{T_2^j}} e^{-\frac{\tilde{\tau}}{T_1^{NV}}} \approx$$
$$e^{-\frac{\tilde{\tau}}{T_2^{NV}}} e^{-\frac{\tilde{\tau}}{T_1^{NV}}} \int_0^R p(\tilde{\tau}, r) dr, \quad (4)$$

where $T_2$ means the nuclear relaxation time and $p_j(\tilde{\tau}, r)$ is the probability of j-th molecule (initially located at the distance r from the center of the detection area) to stay inside the detection volume after time $\tilde{\tau}$ (Figure 3a). The first and second exponents in Equation 4 describe the relaxational decay of the correlation signal due to the finite nuclear $T_2$ time and $T_1$ time of the NV center. The integral part is responsible for the correlation decay due to molecular diffusion and represents the total effect of all the molecules. Integrating over all possible realizations of $p_j(\tilde{\tau})$ for different $D_T$ and using Equation 4, one can find the diffusion coefficient which fits the envelope of the experimental decay curve.

The model which we used to calculate $p_j(\tilde{\tau})$ is presented in Figure 3a. The movement of phospholipids in the membrane can be considered as diffusion of particles due to the Brownian motion[35]. As a result of this process, the hydrogen-containing phospholipid molecules leave the detection volume of an NV center and do not contribute to the phase acquisition any more, leading to the decay of the correlation signal. Therefore, the envelope of the correlation measurement depends on the probability of the molecule (initially located at the distance r from the center of the detection area) to stay inside the detection volume after time $\tilde{\tau}$, p($\tilde{\tau}$, r). If we look at

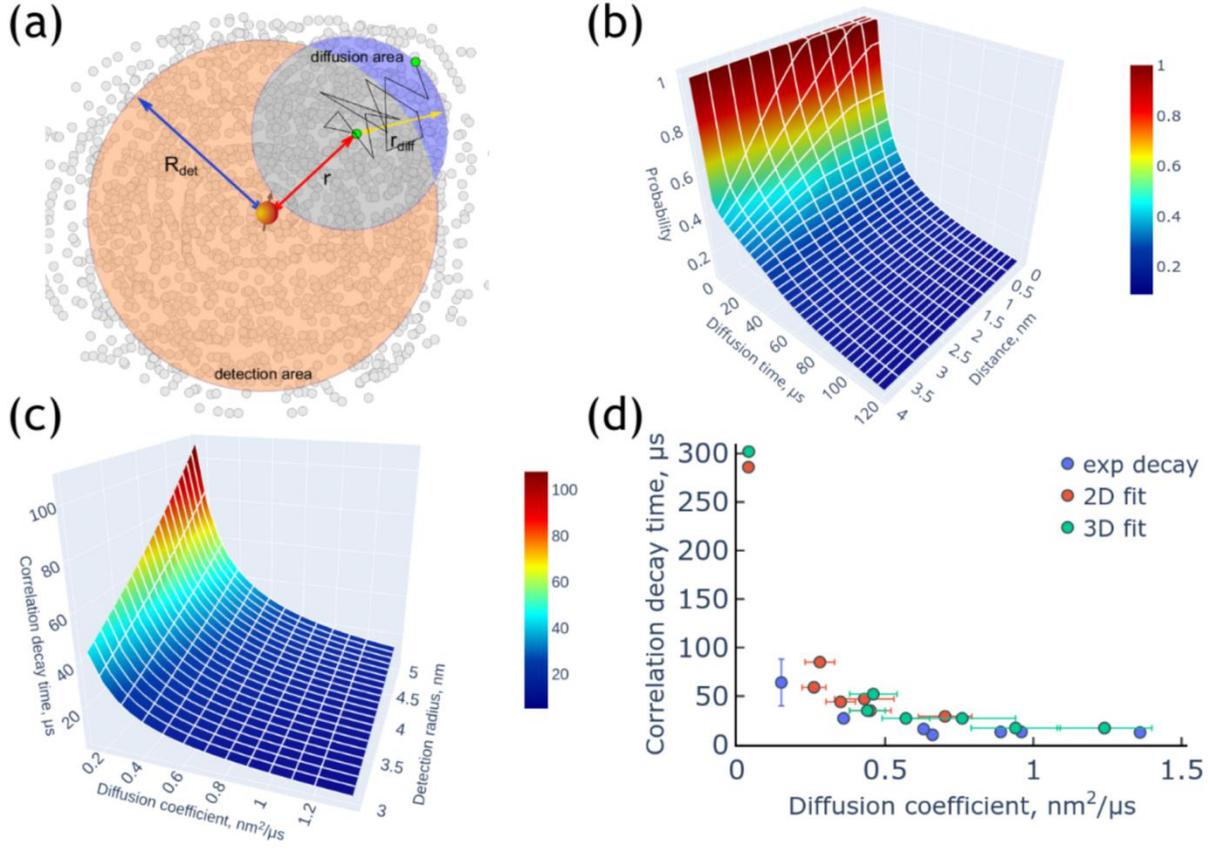

**Figure 3.** Simulations of the system's dynamics. (a) Schematics of the model used to calculate the probability of a molecule to stay inside the detection area for a diffusion time $\tilde{\tau}$. r is the initial distance between the center of the detection area and a molecule. (b) Probability of a molecule to stay within the detection volume as a function of diffusion time $\tilde{\tau}$ and initial distance r from the center of the detection area for $R_{det}$ = 3.93 nm and $D_T$ = 0.358 nm²/μs. (c) Correlation decay time caused only by translational diffusion (without relaxational decay) as a function of $D_T$ and $R_{det}$. (d) Experimental correlation decay times as a function of calculated diffusion coefficients.

Figure 1a from the top, it becomes clear that the detection area is cylindrical, determined by a certain radius $R_{det}$ (Figure 3a), since the detection volume represents a spherical cap and the ¹H signal comes only from the phospholipids. Therefore, p($\tilde{\tau}$, r) can be calculated as the ratio between the area created by the overlapping of the detection area by diffusion area and the diffusion area. Rotational diffusion is at least an order of magnitude slower then translational diffusion and can therefore be neglected[36]. In the framework of this approximation the diffusion area is determined only by the 2D translational diffusion and the diffusion radius over time t is calculated as $r_{diff} = \sqrt{4D_T t}$, where $D_T$ is the coefficient of translational diffusion. Thus, the probability p($\tilde{\tau}$, r) depends on the diffusion coefficient.

For each correlation measurement, we have found $D_T$ which ensured the best approximation of Equation 4 to the envelope of the measured signal. For example, the simulated probability p($\tilde{\tau}$, r) which fits the data from Figure 2c, as a function of the distance from the center of the detection area r and diffusion time $\tilde{\tau}$ is depicted in Figure 3b. This calculation corresponds to the radius of the detection area $R_{det}$ = 3.93 nm and the coefficient of translational diffusion $D_T$ = 0.358 nm²/μs. Then this 3D data was integrated over r (Equation 4) and normalized to get the contribution of the diffusion to the decay of the correlation signal.

Figure 3c shows the decay of the correlation signal due to molecules leaving the detection volume (decay of the integral in Equation 4), where one can see that for more shallow NV centers correlation signal decays faster. This can be essential for the deeper NVs when the diffusion-modulated decay will be longer than the relaxational decay of the correlation measurement. To fit the experimental data, the integral was multiplied by exponential terms, where the nuclear $T_2$ was taken to be 1 ms.

The measured correlation decay times and calculated diffusion coefficients are presented in Figure 3d. For three different diffusion models, the measured coefficients of translational diffusion of phospholipids lie in a range from 0.04 to 1.36 nm$^2$/μs (Table S4). Despite the fact that the exponential model is rather qualitative, it outlines the system's dynamics in a way similar to 2D and 3D cases, i.e. the models are similar good.

The measured broad distribution of diffusion coefficients can be explained by two facts. First, our measurements were done under ambient conditions, i.e. the micelles-containing droplet was not protected from drying out. Indeed, a comparison of this result with literature values for similar phospholipid monolayer[37] shows that the experimentally obtained $D_T$ coefficients correspond to an intermediate case between solid and liquid phase state of the phospholipid monolayer (Figure S4). This argument is also supported by the calculated proton densities based on the known NV depth for each NMR measurement of micelles, which varies from $1.5 \times 10^{28}$ m$^{-3}$ to $2.0 \times 10^{29}$ m$^{-3}$ (Table S2).

Second, any measurement scheme where the monolayer is supported on a solid substrate suffer from surface-induced artefacts[18]. The surface roughness of the substrate has a known impact on the supported lipid bilayers: smoother surface results in the faster diffusion[38]. This also applies to the diamond surface, since its roughness is known to be quite sensitive to the surface treatment[39].

In any case, our findings are rather similar to the recently measured diffusion constant by an ensemble of NV centers using correlation spectroscopy[40]. They observed a change in the diffusion constant from 1.5 nm$^2$/μs to 3.5 nm$^2$/μs when the temperature changes from 26.5 °C to 36.0 °C for liquid sample in an incubator.

In this paper, we have shown that correlation spectroscopy with single shallow NV centers allows studying the diffusion of phospholipids in artificial cell membranes. We used three different diffusion models to fit our experimental data. The experimentally obtained values for the coefficient of translational diffusion lie the range from 0.04 to 1.36 nm$^2$/μs, which corresponds to an intermediate case between solid and liquid phase state of the phospholipid monolayer. NV spectroscopy is thus one of the few methods which could reveal the diffusion heterogeneity on the nanoscale caused by the presence of lipid rafts in the cell membrane, for example.

It is worth noting, that NV based nano-NMR is a non-invasive detection method and does not require the fluorescent labeling. However, there are several limitations which can be imposed on the processes to be studied. First, for sensing with DD based schemes, rather high nuclear spin-density is required. Second, a characteristic timescale of the process should lie within a certain range, determined by the correlation decay time of the sensor (NV center). Partially this problem can be solved by using deeper NVs[41]. At last, the sample must be either chemically attachable to the diamond surface or should at least be in close contact (a couple of nanometers away from the surface).

## ACKNOWLEDGMENTS

The authors acknowledge support from the EU-FET Flagship on Quantum Technologies through the Project ASTERIQS, the European Research Council through the ERC grant SMeL, the DFG through FOR2724, GRK2642 and 2198/2. D.C. acknowledges the support of the Clore foundation and the Clore scholars programme.